\documentclass[conference]{IEEEtran}
\usepackage{cite}
\usepackage{amsmath,amssymb,amsfonts}
\usepackage{algorithmic}
\usepackage{graphicx}
\usepackage{textcomp}
\usepackage{xcolor}
\usepackage{tikz}
\usepackage{subfigure}
\usepackage{arydshln}
\usepackage{pgfplots}

\usetikzlibrary{arrows, automata}
\usetikzlibrary{positioning}
\usetikzlibrary{arrows.meta}
\def\BibTeX{{\rm B\kern-.05em{\sc i\kern-.025em b}\kern-.08em
    T\kern-.1667em\lower.7ex\hbox{E}\kern-.125emX}}

\newenvironment{definition}[1]
{{\bf Def. (#1):}}

\definecolor{_red}{rgb}{1,0,0}
\definecolor{_blue}{rgb}{0,0,1}
\definecolor{_green}{rgb}{0,1,0}
\definecolor{_yellow}{rgb}{1,1,0}

\usepackage{tikz}
\usepackage{textcomp}
\usepackage{hyperref}
\usepackage{lipsum}

\newcommand\copyrighttext{%
  \footnotesize \textcopyright 2018 IEEE. Personal use of this material is permitted.
  Permission from IEEE must be obtained for all other uses, in any current or future
  media, including reprinting/republishing this material for advertising or promotional
  purposes, creating new collective works, for resale or redistribution to servers or
  lists, or reuse of any copyrighted component of this work in other works.\\
  \emph{To appear in the Proceedings of the 2018  IEEE Symposium Series on Computational Intelligence (IEEE SSCI 2018)}}
\newcommand\copyrightnotice{%
\begin{tikzpicture}[remember picture,overlay]
\node[anchor=north,yshift=-5pt] at (current page.north) {\fbox{\parbox{\dimexpr\textwidth-\fboxsep-\fboxrule\relax}{\copyrighttext}}};
\end{tikzpicture}%
}

\begin{document}

\title{Text Summarization as Tree Transduction by Top-Down TreeLSTM
}

\author{\IEEEauthorblockN{Davide Bacciu}
\IEEEauthorblockA{\textit{Department of Computer Science} \\
\textit{University of Pisa}\\
Pisa, Italy \\
bacciu@di.unipi.it}
\and
\IEEEauthorblockN{Antonio Bruno}
\IEEEauthorblockA{\textit{Department of Computer Science} \\
\textit{University of Pisa}\\
Pisa, Italy \\
antonio.bruno@di.unipi.it}
}

\maketitle

\copyrightnotice

\begin{abstract}
Extractive compression is a challenging natural language processing problem. This work contributes by formulating neural extractive compression as a parse tree transduction problem, rather than a sequence transduction task. Motivated by this, we introduce a deep neural model for learning structure-to-substructure tree transductions by extending the standard Long Short-Term Memory, considering the parent-child relationships in the structural recursion.
The proposed model can achieve state of the art performance on sentence compression benchmarks, both in terms of accuracy and compression rate.
\end{abstract}

\begin{IEEEkeywords}
structured-data processing, tree transduction, top-down TreeLSTM, sentence compression
\end{IEEEkeywords}

\section{Introduction}
In the latter years there has been increasing interest in text-to-text rewriting methods for many natural language processing applications, such as machine translation \cite{seq2seq}, question answering \cite{qa} and text compression \cite{text}. In this work, we focus on text compression problems, that is often formulated as  \emph{extractive} compression  \cite{compr}, in which only word deletions are allowed. Given an ordered set of words representing a source sentence 
the aim is to produce the associated compressed text 
by removing any subset of these words, unchanging the order.

The task is typically addressed in the neural network literature by considering it as a sequence-to-sequence transduction problem. Representing natural language sentences as sequences of words has limitations. For instance, it can separate a word from its relevant context (i.e. other relevant words) which, in non-trivial sentences, can be interleaved by a subsequence of words (e.g. a subordinate sentence) irrelevant to the interpretation of the target word. In this paper, we reformulate extractive compression as a structure-to-substructure transduction problem by considering a parse tree representation of the sentence (both original and compressed). This representation  highlights how words are assembled into linguistic substructures according to grammar rules, explicitly capturing the relevant context in which single words should be interpreted. We show how text compression can effectively be solved by learning a structure-to-substructure transduction problem, by extending the standard Long Short-Term Memory (LSTM) to  consider the parent-child relationships of the trees in the structural recursion.

The adaptive processing of tree-structured data is a widely studied machine learning task, starting from the seminal work by \cite{generalframework} which established a general framework for it. Several approaches exists in literature for dealing with tree-structured input samples, while there are considerably less tackling the production of a tree-structured output. Generative models, for instance, have been used to learn a probability distribution on trees exploiting conditional independence relationships induced by the tree structure to simplify inference. According with the direction of tree visit, in literature both top-down \cite{tdthmm} and bottom-up \cite{buthmm,gtmsd,gtmIJCNN} models have been proposed.

Recursive Neural Networks (RecNN) \cite{generalframework} extend the concept of recurrent neural model to handle the processing of tree-structured data. These models typically perform a bottom-up processing of the tree, unfolding the network over the tree structure in such a way that the hidden recursive state of the current tree node depends from that of its children. In practice, this results in an unfolded network resembling a deep multilayer Perceptron with weight sharing. It is easy to understand how such construction is subject to the classical gradient vanishing/explosion  problem \cite{gradVanish}, which has lead to a widespread use of Long Short-Term Memory (LSTM) units  \cite{lstm_} also in the tree-structured domain. The TreeLSTM model in \cite{treelstmbu} has been the first extension of the LSTM cell to handle tree-structures through a bottom-up approach which basically implements a specific instance of the structured data processing framework proposed in the late nineties by \cite{generalframework}. An alternative approach is that put forward in the Tree Echo State Network (TreeESN) \cite{treeesn} where the recursive neurons are randomly initialized according to some dynamic system stability criterion and their weights are not adjusted by the training procedure. Recently, the Hidden Tree Markov Networks (HTNs) \cite{htn} have been proposed as an hybrid approach integrating probabilistic bottom-up models within a neural architecture and learning scheme.

Kernel methods have been widely applied to tree-structured data classification and regression, since they allow a straightforward reuse of kernel-based learning machinery for vectorial data by plugging in an appropriately defined tree kernel.  There has been a large body of research dealing with the definition of efficient and discriminative tree kernels, including syntactic kernels \cite{DBLP:journals/sigkdd/Gartner03}, which compute tree similarity by counting the number of common substructures (e.g. subtrees, paths, etc), as well as generative kernels exploiting an underlying probability distribution over trees, see \cite{bestkernel06} for a recent survey on the topic.

The majority of the works in literature deals with problems of tree classification and regression while the problem of learning structured transductions is seldom addressed. In \cite{IOBUTHMM}, tree transduction has been framed as a problem of learning a conditional generative process between structured samples, providing evidence of how to realize this for isomorph transductions (i.e. where input and output structures have the same skeleton). In this paper, we extend such an approach to deal with a restricted form of non-isomorph transformations, i.e. structure-to-substructure transductions, using a recurrent neural model implementing a top-down visit of the input tree and generating the output tree following the same direction. We show how the proposed approach can be used to effectively deal with the sentence compression problem. Traditionally, extractive compression has been addressed using a sequential representation of the sentences, using both generative/grammar based approaches \cite{crf,CLcorpora} as well as, more recently, through deep recurrent networks \cite{compressLSTM}. Nonetheless, there has been a number of works addressing the problem from a tree-structured perspective, such as in \cite{compr,galley,turner,lapata1,lapata2,LapataCompress}. These, however, use a fully grammar-based approach, where rules are language-depended and hand-engineered and where the only trained part concerns the order of application of the rules.

We propose a Top-Down TreeLSTM (TD-TreeLSTM) that is able to learn structure-to-substructure compression from pairs of parse trees.  We discuss how the top-down  parsing and generation direction allows to capture structural and semantic information in the most effective way for the realization of sentence compression tasks. Contextual information is allowed to flow from the higher, more abstract, levels of the tree reducing ambiguity when deciding if a specific substructure or word needs to be removed from the summary. The TD-TreeLSTM model needs no apriori knowledge concerning the language or grammar, being able to extract the necessary information from the parse tree structure. To the extent of our knowledge, this is also the first use of a top-down recurrent neural model for trees.

\section{Background}

\subsection{Tree-Structured Data and Transductions}

We focus on the problem of learning tree transductions from pairs of input-output trees $(\mathbf{x}^{n},\mathbf{y}^{n})$, where the superscript identifies the $n$-th sample pair in the dataset and it will be omitted when the context is clear. We consider labeled rooted trees where a generic tree is defined by the triplet ${(\mathcal{U}_{n},\mathcal{E}_{n},\mathcal{X}_{n})}$ consisting of a set of nodes $\mathcal{U}_{n} = \{1,\dots,U_n\}$, a set of edges $\mathcal{E}_{n}\subseteq\mathcal{U}_{n}\times\mathcal{U}_{n} $ and a set of labels $\mathcal{X}_{n}$. The term $u \in \mathcal{U}_{n}$ is used to denote a generic tree node, whose direct ancestor, called \emph{parent}, is denoted as $pa(u)$. A node $u$ can have a variable number of direct descendants (\emph{children}), such that the $l$-th child of node $u$ is denoted as $ch_l(u)$. The pair $(u,v) \in \mathcal{E}_{n}$ is used to denote an edge between a generic node and its child and we assume trees to have maximum finite out-degree $L$ (i.e. the maximum number of children of a node). Each node $u$ in the tree is associated with a label $x_u \in \mathcal{X}_{n}$ ($y_u$, for the output tree) which can be of different nature depending on the application, e.g. a vector of continuous-valued features representing word embeddings or a symbol from a discrete and finite alphabet.

A tree transduction defines a mapping from elements of an input domain into element of output domain, where elements from both domains are tree-structured samples. Let $\mathcal{I}^{\#}, \mathcal{O}^{\#}$ denote the input and output domains, respectively,  then a structural transduction is a function $\mathcal{F}: \mathcal{I}^{\#}\rightarrow \mathcal{O}^{\#}$. The transductions considered in this work are based on the concept of tree isomorphism, defined as follows.

\begin{definition}{Tree isomorphism}
Let $\mathbf{x} = {(\mathcal{U},\mathcal{E},\mathcal{X})}$ and $\mathbf{x}' = {(\mathcal{U'},\mathcal{E'},\mathcal{X'})}$, they are isomorphic if exists a bijection\\ $f:\mathcal{U} \rightarrow \mathcal{U'}$ such that $ \forall (u,v)\in \mathcal{E} \iff (f(u), f(v))\in \mathcal{E'}$.
\end{definition}

An equivalent definition can be given using the concept of \emph{skeleton}.

\begin{definition}{Skeleton tree}
Let $\mathbf{x} = {(\mathcal{U},\mathcal{E},\mathcal{X})}$, its skeleton is  $skel(\mathbf{x}) = {(\mathcal{U},\mathcal{E})}$.
\end{definition}

\begin{figure}[h]
\centering
\resizebox{.4\textwidth}{!}
{\includegraphics{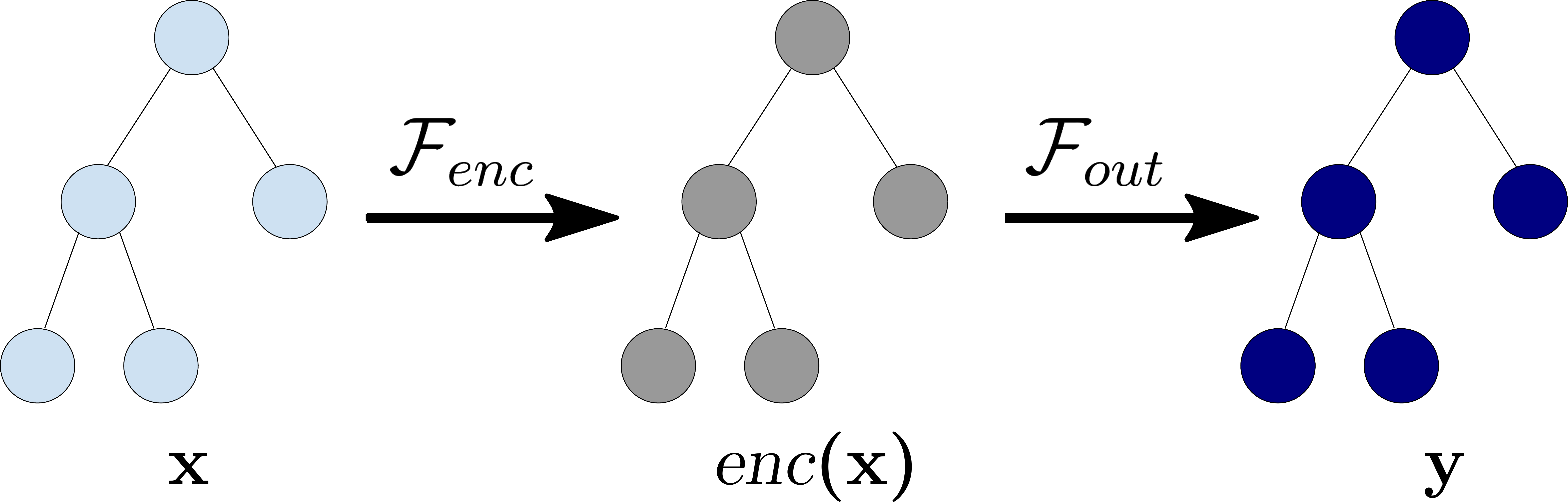}}
\caption{A tree-to-tree transduction. Encoding and output are isomorphic to the input.}
\label{fig:s2s}
\end{figure}

Following such definition, two trees are isomorphic if they have the same skeleton (labels are irrelevant, only structure matters). A general \emph{structure-to-structure} transduction can be realized by an encoding-decoding process where $\mathcal{F} = \mathcal{F}_{out} \circ \mathcal{F}_{enc}$ with:
\begin{equation*}
\mathcal{F}_{enc}: \mathcal{I}^{\#}\rightarrow \mathcal{H}^{\#} \qquad \qquad \mathcal{F}_{out}: \mathcal{H}^{\#}\rightarrow \mathcal{O}^{\#}.
\end{equation*}
The terms $\mathcal{F}_{enc}$ and $\mathcal{F}_{out}$ are the \emph{encoding} and \emph{output} transductions, respectively, and $\mathcal{H}^{\#}$ is the structured state space (e.g. the space of the hidden neuron activations of a recursive neural model). Figure \ref{fig:s2s} shows a tree-to-tree isomorphic transduction in both encoding and output mappings.

\begin{figure}[h]
\centering
\resizebox{.4\textwidth}{!}
{\includegraphics{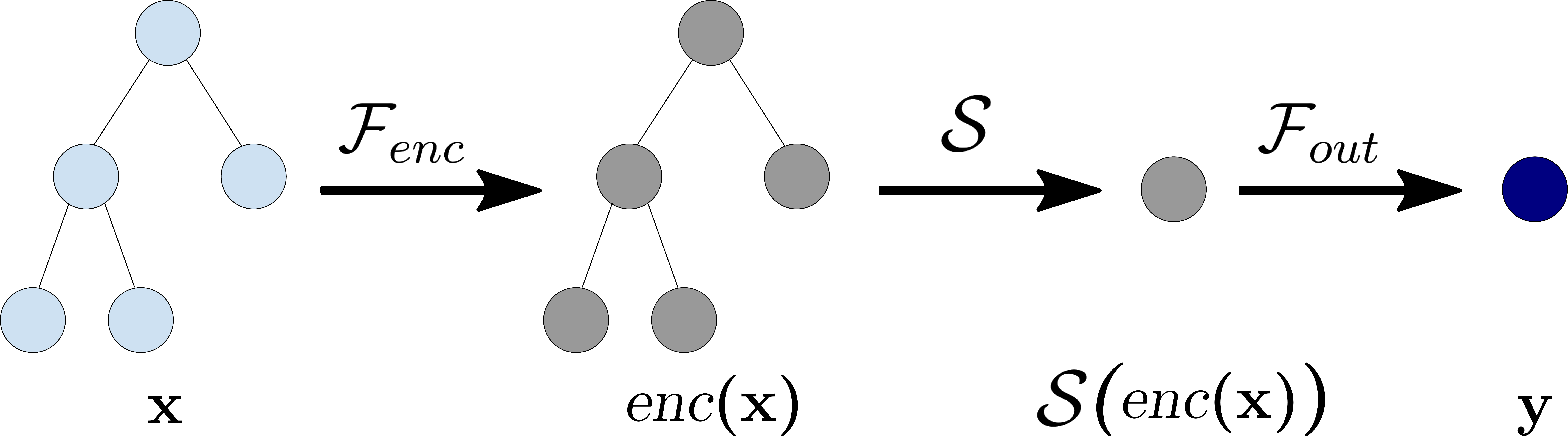}}
\caption{A tree-to-element transduction: the encoding is a structure isomorphic to the input while the output is a vectorial sample.}
\label{fig:s2e}
\end{figure}

Another type of basic transduction is the one shown in Fig. \ref{fig:s2e}, that is a \emph{structure-to-element} or \emph{supersource} transduction. This maps an input tree to a vectorial domain, such as with a tree classification or regression task. In this case the encoding function produces a tree isomorphic to the input one, then a \emph{state mapping function} $\mathcal{S}$ is applied to flatten the encoding structure to a vector which is used by $\mathcal{F}_{out}$ to produce the final output.

In this work, we focus on a particular form of transduction that we call \emph{structure-to-substructure} transduction. We consider the case in which the output tree $\mathbf{y}$ is obtained from the input tree $\mathbf{x}$ by pruning some of its proper subtrees. Conceptually this is a non-isomorphic transduction but, for the purpose of our text compression application, it can be effectively implemented as an isomorphic transduction where the output tree has the labels of the pruned nodes set to a specific \emph{NULL} value.

\subsection{Learning Text Summarization}
Text summarization applications using an extractive compression approach can be coarsely partitioned depending on the sample representation they use.

Sequential models represent the sentence as a sequence of words. In \cite{CLcorpora}, Integer Linear Programming (ILP)  is used to infer globally optimal compression. The text compression problem is modeled by constrained optimization problem using a set of integer variables and associated linear local and global constraints.  The performances of this model are sensitive to the definition of the particular language and grammar dependent constraints, which are  all hand-crafted. In \cite{crf}, compression is casted as a sequence labeling problem using Conditional Random Fields (CRF) in which, for each pair of input-output sequence, the observations $X$ are the words in the input and the random variables $Y$ are the words in the output. The CRF feature functions consider aspects such as word tokens, POS tags, positional features and features extracted from syntactic and discourse parse trees. Again, these are hand engineered aspects and changing language or domain could lead to their redefinition.  A neural based approach is taken by \cite{compressLSTM}, where a LSTM  is used to perform an online/on-the-fly compression of the sequence of input works. In this work, it is
shown that adding an extra input layer to reduce sparsity helps to obtain more compression while performing a pre-training of the model as autoencoder increases the accuracy.

Tree-based models exploit a hierarchical representation of the sentences (both in input and in output) in the attempt to make good use of information about words' contextual relationships. These approaches are typically based on Synchronous Context-Free Grammar (SCFG). Some of them perform discriminative subtree deletion,  such as in \cite{compr}, where a noisy-channel model is used to identify the compression $\mathbf{y}$ which maximizes $P(\mathbf{y})P(\mathbf{x}|\mathbf{y})$ such that $P(\mathbf{y})$ measures the grammatical quality of the compressed sentence and $P(\mathbf{x}|\mathbf{y})$ captures the probability that the source sentence $\mathbf{x}$ is an expansion of the target compression $\mathbf{y}$. This approach has been extended in \cite{galley} using a different lexicalized head-driven Markovianization which yields more robust probability estimates. The authors of \cite{turner} have instead shown that SCFGs are not expressive enough to model structurally complicated compressions, showing how to enhance them through sets of general rules. A different line of research, followed by \cite{lapata1,lapata2,LapataCompress}, perform compression as subtree substitution using another generalization of the SCFG, that is the Synchronous Tree Substitution Grammar (STSG). The fundamental limitation of these grammar-based models is the need to hand-define the transformation rules, which are strictly dependent on the specific task, language and syntactic categories of the parser, which is a key problem for the scalability of the approach.

Our proposed model relaxes the need for such expert information, allowing to infer implicit substitution rules from non-aligned parse trees through the TD-TreeLSTM, whose formalization is given in the following section.

\section{Top-Down Tree Transduction for Text Summarization}
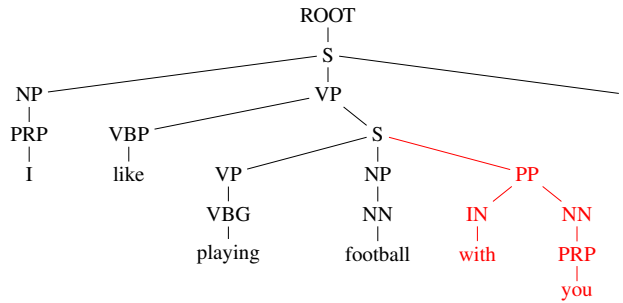
\begin{figure}[h]
\centering
\resizebox{.47\textwidth}{!}
{
\begin{tikzpicture}[node distance=1cm,auto,>=latex']
\tikzstyle{node} = [rectangle, inner sep=0.1cm]
\tikzset{font=\large}

\node[node] (ROOT) at (0,0) {ROOT};

\node[node] (S1) at (0,-1*0.8) {S};

\node[node] (NP1) at (-6,-2*0.8) {NP};
\node[node] (PRP1) at (-6,-3*0.8) {PRP};
\node[node] (I) at (-6,-4*0.8) {I};

\node[node] (dot1) at (6,-2*0.8) {.};
\node[node] (dot2) at (6,-3*0.8) {.};

\node[node] (VP1) at (0,-2*0.8) {VP};
\node[node] (VBP) at (-4,-3*0.8) {VBP};
\node[node] (like) at (-4,-4*0.8) {like};

\node[node] (S2) at (1,-3*0.8) {S};

\node[node] (VP2) at (-2,-4*0.8) {VP};
\node[node] (VBG) at (-2,-5*0.8) {VBG};
\node[node] (playing) at (-2,-6*0.8) {playing};

\node[node] (NP2) at (1,-4*0.8) {NP};
\node[node] (NN1) at (1,-5*0.8) {NN};
\node[node] (football) at (1,-6*0.8) {football};

\node[node] (PP) at (4,-4*0.8) {\textcolor{red}{PP}};

\node[node] (IN) at (3,-5*0.8) {\textcolor{red}{IN}};
\node[node] (with) at (3,-6*0.8) {\textcolor{red}{with}};

\node[node] (NN2) at (5,-5*0.8) {\textcolor{red}{NN}};
\node[node] (PRP2) at (5,-6*0.8) {\textcolor{red}{PRP}};
\node[node] (you) at (5,-7*0.8) {\textcolor{red}{you}};

\draw[-] (ROOT) to (S1);
\draw[-] (S1) to (NP1);
\draw[-] (NP1) to (PRP1);
\draw[-] (PRP1) to (I);
\draw[-] (S1) to (dot1);
\draw[-] (dot1) to (dot2);
\draw[-] (S1) to (VP1);
\draw[-] (VP1) to (VBP);
\draw[-] (VBP) to (like);
\draw[-] (VP1) to (S2);
\draw[-] (S2) to (VP2);
\draw[-] (VP2) to (VBG);
\draw[-] (VBG) to (playing);
\draw[-] (S2) to (NP2);
\draw[-] (NP2) to (NN1);
\draw[-] (NN1) to (football);
\draw[-,red] (S2) to (PP);
\draw[-,red] (PP) to (IN);
\draw[-,red] (IN) to (with);
\draw[-,red] (PP) to (NN2);
\draw[-,red] (NN2) to (PRP2);
\draw[-,red] (PRP2) to (you);

\end{tikzpicture}
}
\caption{An example of a structure-to-substructure transduction between the input parse tree of the sentence ``I like playing football with you." and the substructure representing  its compressed  version ``I like playing football.". The parts of the original tree deleted in the compressed version are marked in red (the corresponding output node labels are set to \emph{NULL}).}
\label{fig:example}
\end{figure}

We propose a model to realize text compression as structure-to-substructure transductions. It receives in input a parse tree (i.e. costituency tree) representation of the input text and returns in output a parse tree representing the compressed sentence (Fig. \ref{fig:example}). We show how this transduction process can be learned by a TD-TreeLSTM unfolding on all the root-to-leaf paths of the parse trees, whose internal nodes represent syntactic categories, while the leaves are labeled by dictionary words.

\subsection{Top-Down TreeLSTM (TD-TreeLSTM)}
Several works have been focused on extending Recurrent Neural Networks (RNN) to deal with tree structured data. Lately, most of these works focused on tree-structured extensions of LSTM cells and networks. The primary source of differentiation between the different approaches concerns the processing direction.
\begin{LaTeXdescription}
\item[Bottom Up (BU):] tree processing flows from leaves to the root. These are recursive models \cite{treelstmbu} mainly used as one-pass encoder of the structure, following the assumption that a node hidden state computed recursively from its children states is a ``good" vectorial summary of the information in all the subtree rooted in the node.
\item[Top Down (TD):] tree processing flows from the root to the leaves. LSTM models belonging to this category are mainly used in generative settings, where one wants to generate the children of a node based on the hidden state of the parent \cite{topdowndep}\cite{drnn}. Their use as encoders of the full structure is not common, as this would require some form of mapping function summarizing the whole tree into a single encoding vector (e.g. the mean of the hidden states of all the nodes in the tree).
\end{LaTeXdescription}

\begin{figure}[h]
\centering
\resizebox{.485\textwidth}{!}
{
\begin{tikzpicture}[node distance=1cm,auto,>=latex', scale = 0.65, transform shape]
\tikzstyle{op} = [draw,circle, inner sep=0cm]
\tikzstyle{fun} = [draw,circle, minimum size = 1cm, inner sep = 0.05cm]
\tikzstyle{dot} = [draw,shape=circle,minimum size=0.2cm,inner sep=0, fill = black]
\tikzstyle{fake} = [coordinate]
\tikzstyle{dritto} = [-{Latex[length=2mm]}]
\tikzstyle{piatto} = [-]
\tikzstyle{curvasu} = [dritto,out=0,in=-90]
\tikzstyle{curvagiu} = [dritto,out=0,in=90]
\tikzset{font=\large}

 \foreach \place/\name/\label in {{(0,0)/tanh1/$\tanh$}, {(0,-2)/sigmi/$\sigma$}, {(0,-4)/sigmf/$\sigma$}, {(0,-6)/sigmo/$\sigma$},{(6,-2)/tanh2/$\tanh$}} \node[fun] (\name) at \place {\label};

 \foreach \place/\name/\label in {{(3,-2)/piu/$+$},{(1.5,-1.0)/x1/$\times$},{(3,-4)/x2/$\times$},{(8,-2)/x3/$\times$}} \node[op] (\name) at \place {\label};

\node[dot] (dot3) at (4.5,-2) {};
\node[fake] (fake1) at (-2.7,-7) {};
\node[fake] (fake2) at (2,-7) {};
\node[fake] (dot2) at (-3.5,-4) {};

\node[dot] (hub) at (-2.5,-3) {};

\node[fake] (ul) at (-3.2,1) {};
\node[fake] (ur) at (9.8,1) {};
\node[fake] (dl) at (-3.2,-6.8) {};
\node[fake] (dr) at (9.8,-6.8) {};
\node[fake] (dot99) at (-5,-5) {};
\node[fake] (out) at (10,-2) {};
\node[fake] (output) at (11,-2) {};

\node[fake] (input) at (-5,-3) {};
\node[fake] (inputint) at (-3.5,-3) {};

\node[fake] (output2) at (11,-.5) {};
\node[fake] (fakeX) at (6,-.5){};
\node[fake] (fakeY) at (2.3,-5){};
\node[fake] (input2) at (-5,-5) {};
\node[fake] (inputint2) at (-3.5,-5) {};

\foreach \inizio/\fine/\posi/\posf/\label/\posl in {{tanh1/x1/east/north/$r_u$/above},{x1/piu/east/north//above}} \draw [curvagiu] (\inizio.\posi) to node[\posl]{\label} (\fine.\posf) ;

\foreach \inizio/\fine/\posi/\posf/\label/\posl in {{sigmi/x1/east/south/$i_u$/below},{sigmo/x3/east/south/$o_u$/above}} \draw [curvasu] (\inizio.\posi) to node[\posl]{\label} (\fine.\posf) ;

\foreach \inizio/\fine/\label/\posl in {{out/output//}, {dot3/tanh2//}, {sigmf/x2/$f_u$/}, {x2/piu//}, {tanh2/x3//}} \draw [dritto] (\inizio) to node [\posl] {\label} (\fine);

\foreach \inizio/\fine/\label/\posl in {{x3/out/$h_u$/}, {inputint/hub//}} \draw [piatto] (\inizio) to node [\posl] {\label} (\fine);

\draw[-] (ul) to (ur);
\draw[-] (ul) to (dl);
\draw[-] (ur) to (dr);
\draw[-] (dl) to (dr);

\draw [-] (input) to node{$x_u,h_{pa(u)}$} (inputint) ;
\draw [-] (input2) to node{$c_{pa(u)}$} (inputint2) ;
\draw[-] (inputint2) to node{} (fakeY);


\draw[dritto, out=90, in=180] (hub.north) to (tanh1);
\draw[dritto, out=45, in=180] (hub) to (sigmi);
\draw[dritto, out=-45, in=180] (hub) to (sigmf);
\draw[dritto, out=-90, in=180] (hub.south) to (sigmo);
\draw[-, out=90, in=180] (dot3.north) to node{} (fakeX);
\draw[dritto] (fakeX) to node{} (output2);
\draw[-] (piu) to node {$c_u$} (dot3);
\draw[curvasu] (fakeY) to node{} (x2);

\end{tikzpicture}
}
\caption{Graphical representation of the TD-TreeLSTM unit: it receives in input the current node label, the parent's hidden and memory cell states used to compute its hidden and memory cell states which are propagated to the children.}
\label{fig:lstmcell}
\end{figure}
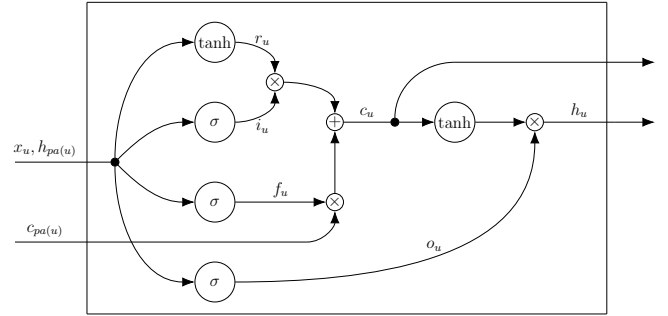

Here, we put forward the idea that a TD recurrent model provides an effective means to realize structure-to-substructure transductions. The underlying neural machinery of our proposed model is an LSTM cell, hence we call our approach TD-TreeLSTM. To the extent of our knowledge, this is the first application of a top-down LSTM model in the context of structure-to-substructure transductions. The TD-TreeLSTM unit activation for a generic node $u$ is formally defined by the following equations (see Fig. \ref{fig:lstmcell} for the corresponding graphical representation)
\begin{align}
 r_u &= \tanh \biggl( W^{(r)} x_u + U^{(r)} h_{pa(u)} + b^{(r)} \biggr) \tag{standard recurrence}
\\
 i_u &= \sigma \biggl( W^{(i)} x_u + U^{(i)} h_{pa(u)} + b^{(i)} \biggr) \tag{input gate}
\\
 o_u &= \sigma \biggl( W^{(o)} x_u + U^{(o)} h_{pa(u)} + b^{(o)} \biggr) \tag{output gate}
\\
 f_u &= \sigma \biggl( W^{(f) }x_u + U^{(f)} h_{pa(u)} + b^{(f)} \biggr) \tag{forget gate}
\\\
 c_u &= i_u \odot r_u+ f_u \odot c_{pa(u)} \tag{memory cell state}
\\
 h_u &= o_u \odot \tanh (c_u) \tag{hidden state}
\end{align}
\noindent with the term $x_u$ denoting unit input (i.e. a vectorial representation of the node label), $h_{pa(u)}$ and $c_{pa(u)}$ are, respectively, the hidden state and the memory cell state of the node's parent, $\sigma$ is the sigmoid activation function and $\odot$ is elementwise multiplication. It can be seen that the formal model of this unit is that of a standard LSTM unit for sequences; however, this will be TD unfolded over the tree by following in parallel all the root to leaves paths.

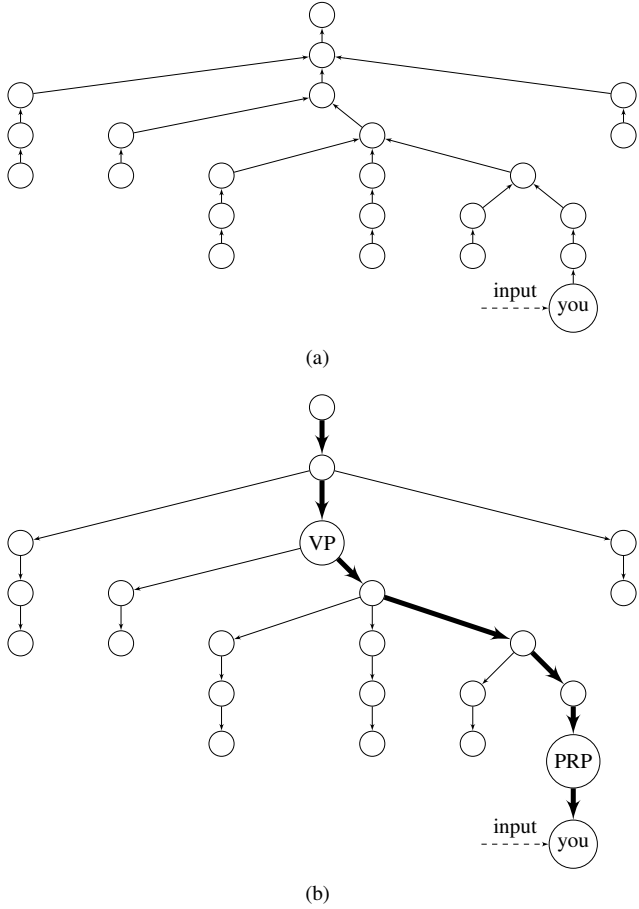
\begin{figure}
\centering
\subfigure[\label{fig:bottomup}]
{
\resizebox{.47\textwidth}{!}
{
\begin{tikzpicture}[node distance=1cm,auto,>=latex']
\tikzstyle{node} = [circle,draw, minimum size = .5cm, inner sep=0.1cm]
\tikzset{font=\large}

\node[node] (ROOT) at (0,0) {};

\node[node] (S1) at (0,-1*0.8) {};

\node[node] (NP1) at (-6,-2*0.8) {};
\node[node] (PRP1) at (-6,-3*0.8) {};
\node[node] (I) at (-6,-4*0.8) {};

\node[node] (dot1) at (6,-2*0.8) {};
\node[node] (dot2) at (6,-3*0.8) {};

\node[node] (VP1) at (0,-2*0.8) {};
\node[node] (VBP) at (-4,-3*0.8) {};
\node[node] (like) at (-4,-4*0.8) {};

\node[node] (S2) at (1,-3*0.8) {};

\node[node] (VP2) at (-2,-4*0.8) {};
\node[node] (VBG) at (-2,-5*0.8) {};
\node[node] (playing) at (-2,-6*0.8) {};

\node[node] (NP2) at (1,-4*0.8) {};
\node[node] (NN1) at (1,-5*0.8) {};
\node[node] (football) at (1,-6*0.8) {};

\node[node] (PP) at (4,-4*0.8) {};

\node[node] (IN) at (3,-5*0.8) {};
\node[node] (with) at (3,-6*0.8) {};

\node[node] (NN2) at (5,-5*0.8) {};
\node[node] (PRP2) at (5,-6*0.8) {};
\node[node] (you) at (5,-7.3*0.8) {you};

\node[circle] (youin) at (3,-7.3*0.8) {};

\draw[<-] (ROOT) to (S1);
\draw[<-] (S1) to (NP1);
\draw[<-] (NP1) to (PRP1);
\draw[<-] (PRP1) to (I);
\draw[<-] (S1) to (dot1);
\draw[<-] (dot1) to (dot2);
\draw[<-] (S1) to (VP1);
\draw[<-] (VP1) to (VBP);
\draw[<-] (VBP) to (like);
\draw[<-] (VP1) to (S2);
\draw[<-] (S2) to (VP2);
\draw[<-] (VP2) to (VBG);
\draw[<-] (VBG) to (playing);
\draw[<-] (S2) to (NP2);
\draw[<-] (NP2) to (NN1);
\draw[<-] (NN1) to (football);
\draw[<-] (S2) to (PP);
\draw[<-] (PP) to (IN);
\draw[<-] (IN) to (with);
\draw[<-] (PP) to (NN2);
\draw[<-] (NN2) to (PRP2);
\draw[<-] (PRP2) to (you);
\draw[->,dashed] (youin) tonode{input} (you);

\end{tikzpicture}
}
}
\vspace{3mm}
\subfigure[\label{fig:topdown}]
{
\resizebox{.47\textwidth}{!}
{
\begin{tikzpicture}[node distance=1cm,auto,>=latex']
\tikzstyle{node} = [circle,draw, minimum size = .5cm, inner sep=0.1cm]
\tikzset{font=\large}

\node[node] (ROOT) at (0,0.7) {};

\node[node] (S1) at (0,-0.5) {};

\node[node] (NP1) at (-6,-2) {};
\node[node] (PRP1) at (-6,-3) {};
\node[node] (I) at (-6,-4) {};

\node[node] (dot1) at (6,-2) {};
\node[node] (dot2) at (6,-3) {};

\node[node] (VP1) at (0,-2) {VP};
\node[node] (VBP) at (-4,-3) {};
\node[node] (like) at (-4,-4) {};

\node[node] (S2) at (1,-3) {};

\node[node] (VP2) at (-2,-4) {};
\node[node] (VBG) at (-2,-5) {};
\node[node] (playing) at (-2,-6) {};

\node[node] (NP2) at (1,-4) {};
\node[node] (NN1) at (1,-5) {};
\node[node] (football) at (1,-6) {};

\node[node] (PP) at (4,-4) {};

\node[node] (IN) at (3,-5) {};
\node[node] (with) at (3,-6) {};

\node[node] (NN2) at (5,-5) {};
\node[node] (PRP2) at (5,-6.35) {PRP};
\node[node] (you) at (5,-8) {you};

\node[circle] (youin) at (3,-8) {};

\draw[->,line width=1mm] (ROOT) to (S1);
\draw[->] (S1) to (NP1);
\draw[->] (NP1) to (PRP1);
\draw[->] (PRP1) to (I);
\draw[->] (S1) to (dot1);
\draw[->] (dot1) to (dot2);
\draw[->,line width=1mm] (S1) to (VP1);
\draw[->] (VP1) to (VBP);
\draw[->] (VBP) to (like);
\draw[->,line width=1mm] (VP1) to (S2);
\draw[->] (S2) to (VP2);
\draw[->] (VP2) to (VBG);
\draw[->] (VBG) to (playing);
\draw[->] (S2) to (NP2);
\draw[->] (NP2) to (NN1);
\draw[->] (NN1) to (football);
\draw[->,line width=1mm] (S2) to (PP);
\draw[->] (PP) to (IN);
\draw[->] (IN) to (with);
\draw[->,line width=1mm] (PP) to (NN2);
\draw[->,line width=1mm] (NN2) to (PRP2);
\draw[->,line width=1mm] (PRP2) to (you);
\draw[->,dashed] (youin) tonode{input} (you);

\end{tikzpicture}
}
}
\caption{Bottom-up (a) vs top-down (b) approach on the parse tree in Fig. \ref{fig:example}. The arrows denote the direction of the tree processing flow. Note how the top-down approach allows to condition the encoding of the leaf labeled with ``you" based on the context given by node ``PRP" and, recursively, by its ancestors (e.g. ``VP").}
\label{fig:tdvsbu}
\end{figure}

The rationale behind the choice of a TD approach is founded on the specificity of the tree-compression application. In a parse tree the dictionary words occur only as leaf labels; a disambiguation of their interpretation can be performed by considering their context which can only come from their parents, given that they have no children by definition. Hence, it follows that a parent-to-children information flow is more relevant than a children-to-parent one to determine if a word, represented necessarily by a leaf node, has to be included or not in a summary. Let us consider a trivial example: the word ``you" could be interpreted either as a subject or complement, depending on the context. It is highly likely that, in the former case, the word needs to be kept during compression. On the contrary, it is highly likely that it will be deleted in the latter interpretation. Clearly, a purely BU model cannot distinguish between the two cases.  The example above is rendered graphically in Fig. \ref{fig:tdvsbu}: in the BU case, shown in Fig. \ref{fig:bottomup}, the state (and the corresponding output) of the leaf node ``you" is conditioned only on its input label and its encoding cannot change depending on where the ``you" labeled node occur in the structure. Conversely, the TD case depicted in Fig. \ref{fig:topdown} shows how the state of the ``you" node is conditioned on the encoding of its parent and recursively from the path that has lead to the node. Therefore the occurrence of word ``you" in different parts of the tree can be associated to different encodings of the corresponding leaf depending on the context dictated by the path leading to it.

\subsection{Output Representation}
The previous subsection has described how to obtain a state encoding for the tree nodes using a TD context propagation. To realize the transduction, we need to define also an output function generating the predicted label for each node (which includes the special \emph{NULL} label to denote the absence of a node in the output structure).
In the following, we define two alternative approaches tailored to the text summarization application. Note that, for this specific application, we can restrict to generating output labels only for the leaves (which are the only nodes where words can appear).

\subsubsection{Vectorial Output Layer}\label{sec:vectorial}
node output is word encoded using the same dictionary of the input tree labels. These are typically vector encoded using either a one-hot or a word embedding representation. The input dictionary is augmented with a special \emph{NULL} vector (Fig. \ref{fig:weout}). Formally, the output for a generic node $u$ is computed by
\begin{align}
\overline{out}_u &= \overline{W}_u^{(h)} h_u + \overline{W}_u^{(c)} c_u + \overline{b} \tag{vectorial output}
\end{align}
with $h_u$ and $c_u$ being the hidden and memory cell states, respectively. To associate a specific word of the dictionary to the output vector generated by the equation above, we use a simple nearest neighbor approach, where we associate $\overline{out}_u$ with the closest vector in the words dictionary. Given this interpretation of the output function, we can associate it with a \emph{Mean Squared Error} (MSE) loss to drive the training phase optimization, by minimizing the discrepancy between the ground-truth outputs (i.e. the words in the summary text) and the output computed by the model for each node.

\subsubsection{Binary Output Layer}
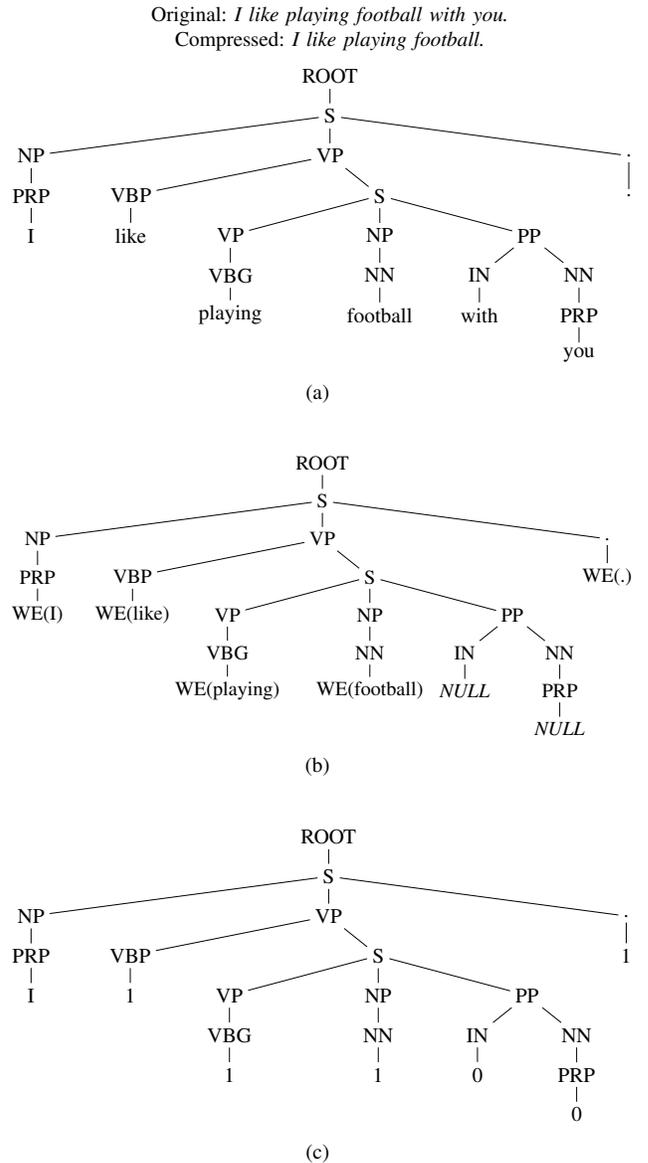
\begin{figure}
\centering
\subfigure[\label{fig:parsetreesent}]
{
\resizebox{.47\textwidth}{!}
{
\begin{tikzpicture}[node distance=1cm,auto,>=latex']
\tikzstyle{node} = [rectangle, inner sep=0.1cm]
\tikzset{font=\large}

\node[node] (Orig) at (0,1.2) {Original: \emph{I like playing football with you.}};
\node[node] (Compressed) at (0,.7) {Compressed: \emph{I like playing football.}};

\node[node] (ROOT) at (0,0) {ROOT};

\node[node] (S1) at (0,-1*0.8) {S};

\node[node] (NP1) at (-6,-2*0.8) {NP};
\node[node] (PRP1) at (-6,-3*0.8) {PRP};
\node[node] (I) at (-6,-4*0.8) {I};

\node[node] (dot1) at (6,-2*0.8) {.};
\node[node] (dot2) at (6,-3*0.8) {.};

\node[node] (VP1) at (0,-2*0.8) {VP};
\node[node] (VBP) at (-4,-3*0.8) {VBP};
\node[node] (like) at (-4,-4*0.8) {like};

\node[node] (S2) at (1,-3*0.8) {S};

\node[node] (VP2) at (-2,-4*0.8) {VP};
\node[node] (VBG) at (-2,-5*0.8) {VBG};
\node[node] (playing) at (-2,-6*0.8) {playing};

\node[node] (NP2) at (1,-4*0.8) {NP};
\node[node] (NN1) at (1,-5*0.8) {NN};
\node[node] (football) at (1,-6*0.8) {football};

\node[node] (PP) at (4,-4*0.8) {PP};

\node[node] (IN) at (3,-5*0.8) {IN};
\node[node] (with) at (3,-6*0.8) {with};

\node[node] (NN2) at (5,-5*0.8) {NN};
\node[node] (PRP2) at (5,-6*0.8) {PRP};
\node[node] (you) at (5,-7*0.8) {you};

\draw[-] (ROOT) to (S1);
\draw[-] (S1) to (NP1);
\draw[-] (NP1) to (PRP1);
\draw[-] (PRP1) to (I);
\draw[-] (S1) to (dot1);
\draw[-] (dot1) to (dot2);
\draw[-] (S1) to (VP1);
\draw[-] (VP1) to (VBP);
\draw[-] (VBP) to (like);
\draw[-] (VP1) to (S2);
\draw[-] (S2) to (VP2);
\draw[-] (VP2) to (VBG);
\draw[-] (VBG) to (playing);
\draw[-] (S2) to (NP2);
\draw[-] (NP2) to (NN1);
\draw[-] (NN1) to (football);
\draw[-] (S2) to (PP);
\draw[-] (PP) to (IN);
\draw[-] (IN) to (with);
\draw[-] (PP) to (NN2);
\draw[-] (NN2) to (PRP2);
\draw[-] (PRP2) to (you);

\end{tikzpicture}
}
}

\vspace{3mm}
\subfigure[\label{fig:weout}]
{
\resizebox{.47\textwidth}{!}
{
\begin{tikzpicture}[node distance=1cm,auto,>=latex']
\tikzstyle{node} = [rectangle, inner sep=0.1cm]
\tikzset{font=\large}

\node[node] (ROOT) at (0,0) {ROOT};

\node[node] (S1) at (0,-1*0.8) {S};

\node[node] (NP1) at (-6,-2*0.8) {NP};
\node[node] (PRP1) at (-6,-3*0.8) {PRP};
\node[node] (I) at (-6,-4*0.8) {WE(I)};

\node[node] (dot1) at (6,-2*0.8) {.};
\node[node] (dot2) at (6,-3*0.8) {WE(.)};

\node[node] (VP1) at (0,-2*0.8) {VP};
\node[node] (VBP) at (-4,-3*0.8) {VBP};
\node[node] (like) at (-4,-4*0.8) {WE(like)};

\node[node] (S2) at (1,-3*0.8) {S};

\node[node] (VP2) at (-2,-4*0.8) {VP};
\node[node] (VBG) at (-2,-5*0.8) {VBG};
\node[node] (playing) at (-2,-6*0.8) {WE(playing)};

\node[node] (NP2) at (1,-4*0.8) {NP};
\node[node] (NN1) at (1,-5*0.8) {NN};
\node[node] (football) at (1,-6*0.8) {WE(football)};

\node[node] (PP) at (4,-4*0.8) {PP};

\node[node] (IN) at (3,-5*0.8) {IN};
\node[node] (with) at (3,-6*0.8) {\emph{NULL}};

\node[node] (NN2) at (5,-5*0.8) {NN};
\node[node] (PRP2) at (5,-6*0.8) {PRP};
\node[node] (you) at (5,-7*0.8) {\emph{NULL}};

\draw[-] (ROOT) to (S1);
\draw[-] (S1) to (NP1);
\draw[-] (NP1) to (PRP1);
\draw[-] (PRP1) to (I);
\draw[-] (S1) to (dot1);
\draw[-] (dot1) to (dot2);
\draw[-] (S1) to (VP1);
\draw[-] (VP1) to (VBP);
\draw[-] (VBP) to (like);
\draw[-] (VP1) to (S2);
\draw[-] (S2) to (VP2);
\draw[-] (VP2) to (VBG);
\draw[-] (VBG) to (playing);
\draw[-] (S2) to (NP2);
\draw[-] (NP2) to (NN1);
\draw[-] (NN1) to (football);
\draw[-] (S2) to (PP);
\draw[-] (PP) to (IN);
\draw[-] (IN) to (with);
\draw[-] (PP) to (NN2);
\draw[-] (NN2) to (PRP2);
\draw[-] (PRP2) to (you);

\end{tikzpicture}
}
}

\vspace{3mm}
\subfigure[\label{fig:binout}]
{
\resizebox{.47\textwidth}{!}
{
\begin{tikzpicture}[node distance=1cm,auto,>=latex']
\tikzstyle{node} = [rectangle, inner sep=0.1cm]
\tikzset{font=\large}

\node[node] (ROOT) at (0,0) {ROOT};

\node[node] (S1) at (0,-1*0.8) {S};

\node[node] (NP1) at (-6,-2*0.8) {NP};
\node[node] (PRP1) at (-6,-3*0.8) {PRP};
\node[node] (I) at (-6,-4*0.8) {I};

\node[node] (dot1) at (6,-2*0.8) {.};
\node[node] (dot2) at (6,-3*0.8) {1};

\node[node] (VP1) at (0,-2*0.8) {VP};
\node[node] (VBP) at (-4,-3*0.8) {VBP};
\node[node] (like) at (-4,-4*0.8) {1};

\node[node] (S2) at (1,-3*0.8) {S};

\node[node] (VP2) at (-2,-4*0.8) {VP};
\node[node] (VBG) at (-2,-5*0.8) {VBG};
\node[node] (playing) at (-2,-6*0.8) {1};

\node[node] (NP2) at (1,-4*0.8) {NP};
\node[node] (NN1) at (1,-5*0.8) {NN};
\node[node] (football) at (1,-6*0.8) {1};

\node[node] (PP) at (4,-4*0.8) {PP};

\node[node] (IN) at (3,-5*0.8) {IN};
\node[node] (with) at (3,-6*0.8) {0};

\node[node] (NN2) at (5,-5*0.8) {NN};
\node[node] (PRP2) at (5,-6*0.8) {PRP};
\node[node] (you) at (5,-7*0.8) {0};

\draw[-] (ROOT) to (S1);
\draw[-] (S1) to (NP1);
\draw[-] (NP1) to (PRP1);
\draw[-] (PRP1) to (I);
\draw[-] (S1) to (dot1);
\draw[-] (dot1) to (dot2);
\draw[-] (S1) to (VP1);
\draw[-] (VP1) to (VBP);
\draw[-] (VBP) to (like);
\draw[-] (VP1) to (S2);
\draw[-] (S2) to (VP2);
\draw[-] (VP2) to (VBG);
\draw[-] (VBG) to (playing);
\draw[-] (S2) to (NP2);
\draw[-] (NP2) to (NN1);
\draw[-] (NN1) to (football);
\draw[-] (S2) to (PP);
\draw[-] (PP) to (IN);
\draw[-] (IN) to (with);
\draw[-] (PP) to (NN2);
\draw[-] (NN2) to (PRP2);
\draw[-] (PRP2) to (you);

\end{tikzpicture}
}
}
\caption{Different output representations for the same compressed tree: (a) shows an example sentence, its parse tree and compression; (b) shows the representation of the compressed tree using a vectorial output (WE($w$) denotes the word embedding of $w$) and (c) shows the corresponding binary output representation.}
\label{fig:outs}
\end{figure}

node output is a binary value. Given a node $u$, its output will be equal to $1$ if the node will be present (together with its label) in the output tree and it will be equal to $0$ otherwise (Fig. \ref{fig:binout}). The corresponding output rule is
\begin{align}
\widehat{out}_u &= \sigma \biggl(\widehat{W}_u^{(h)} h_u + \widehat{W}_u^{(c)} c_u + \widehat{b} \biggr) \tag{binary output}
\end{align}
with $h_u$ and $c_u$ being the hidden and memory cell states, respectively. The term $\sigma$ denotes the sigmoid activation function used to produce an output value representing the probability of node presence: the value of $\widehat{out}_u$ is confronted to a threshold (fixed to $0.5$) to ultimately determine if the node is present or not in the output tree. The training loss associated to this output function is \emph{Binary Cross Entropy} (BCE).

\section{Experiments and Results}
In the following we confront the empirical performance of the proposed TD-TreeLSTM model, on sentence compression tasks, with that of two state-of-the-art approaches: a LSTM-based model \cite{compressLSTM} using a sequential representation of the sentence and a generative model \cite{LapataCompress} based on a tree-structured representation of the text.

\subsection{Experimental Setup}
The experiments are based on the CLWritten and CLSpoken corpora \cite{CLcorpora}, which are widely used for evaluating sentence compression techniques. The former was created by sampling from written sources, while the latter was created by manually transcribed speech of broadcast news stories. Since CLSpoken is created from a speech corpus, it often contains incomplete and ungrammatical utterances and speech artifacts such as disfluencies, false starts and hesitations. The corresponding shortened/compressed form of all sentences were created manually. Here, the annotators were asked to produce the smallest possible target compression by deleting extraneous words from the source, without changing the word order and the meaning of the sentences. In particular, three annotators have been used for the CLSpoken benchmark: we have considered each annotator independently providing then the overall CLSpoken performance as the average on the three annotators.

The corpora provide sentences in sequential form (see \cite{CLcorpora} for more details). The corresponding tree representation has been obtained using the (constituency) \emph{Stanford Parser} \cite{stanfordparser} on the original sentences. We will use the term POStag, from now on, to refer to the categories associated to the internal nodes of the parse trees.

Node labels can be of two kinds depending on the node type: vocabulary words are represented through \emph{word embeddings}, obtained using the \emph{word2vec} \cite{word2vec} approach by performing $1000$ training epochs with negative sampling method; POStag categories, instead, are represented using a one-hot encoding. The statistics of the label representations for the two corpora are reported in Table \ref{tab:sets}.
\begin{table}[h]
\newcommand\T{\rule{0pt}{2.5ex}}       
\caption{Input Label Info - Vector Size is the dimension of the input vectors, Set Size is the number of different input values for each type of label, Vector Type is the encoding used for the input labels}
\begin{center}
\resizebox{.45\textwidth}{!}{
\begin{tabular}{l|c|c|c|c}
\multicolumn{1}{c|}{\begin{tabular}{c}{}\\{\bf Property}\end{tabular}} & \multicolumn{2}{c|}{\bf{POStag Set}} & \multicolumn{2}{c}{\bf{Word Set}}\\
 & {\bf\emph{CLWritten}} & {\bf\emph{CLSpoken}} & {\bf\emph{CLWritten}} & {\bf\emph{CLSpoken}}\\
\hline \T
Vector Size &  68 & 66 & 68 & 66 \\
Set Size &  68 & 66 & 8379 & 4272 \\
Vector Type & \multicolumn{2}{c|}{one-hot} &  \multicolumn{2}{c}{real-valued} \\
\end{tabular}
}
\label{tab:sets}
\end{center}
\end{table}

For model selection and validation purposes we have split each corpus in training set, validation and test sets, along the lines of the experimental setups defined for the baseline models in \cite{LapataCompress}\cite{compressLSTM}. In particular, the partition of the corpora is performed using a hold-out approach with the same proportion of training/validation/test used in \cite{CLcorpora}, i.e. 903/63/882 for CLWritten and 882/78/410 for CLSpoken.

Performance on the two corpora has been assessed using the metrics used in the case of the two baseline models:
\begin{LaTeXdescription}
\item[accuracy or importance factor:] this metric measure how much of the important information is retained. Accuracy is evaluated using \emph{Simple String Accuracy} (SSA) \cite{ssa}, which is based on the string edit distance between the compressed output generated by the model and the reference ground-truth compression;
\item[compression rate:] this metric measures how much of the original sentence remains after compression. Compression rate is defined as the length of the compressed sentence divided by the original length, so lower values indicate more compression;
\item[F1 score:] provides an assessment similar to the accuracy defined above.
\end{LaTeXdescription}

Model selection and \emph{early stopping} decisions are taken by considering an hybrid metric, computed on the validation set, which trades off accuracy and compression according to the following definition
\[
t = \frac{\text{accuracy}^2}{\text{compression rate}}.
\]
Training has been performed by \emph{BackPropagation Through Structure} (BPTS) \cite{bpts} in combination with an Adam optimizer \cite{adam}. $L_{2}$ penalization term has been added to the loss function for the sake of model regularization, using a penalization weight fixed to $\lambda = 10^{-4}$. Model selection has been performed on the number of LSTM units which, after a preliminary coarse grained search, has been selected from the following set of values: 200, 250, 300, 350, 400. Additionally, we have validated the choice of the type of output layer: when considering the vectorial output representation, we have set the  \emph{NULL} element to be the vector with all entries equal to $1$  in order to use the cosine distance as vector-word similarity.

\subsection{Results}
Validation performances for the best TD-TreeLSTM configurations are summarized in Table \ref{tab:validationbin} and \ref{tab:validationwe} for the binary and vectorial outputs, respectively. Overall the binary output seems more effective in optimizing the trade-off metric $t$, in particular because of a more effective compression.
\begin{table}[hb]
\newcommand\T{\rule{0pt}{2.5ex}}       
\caption{Validation performance for TD-TreeLSTM with Binary Output}
\begin{center}
\resizebox{.47\textwidth}{!}
{
\begin{tabular}{l|c|c|c|c}
\multicolumn{1}{c|}{\bf{Corpus}} & \bf{Memory Size} & \bf{Accuracy \%} & \bf{Compress. (\emph{Gold}) \%} & \bf{t}\\
\hline \T
CLWritten & 250 &  74.57 & 72.11 (\emph{73.21}) & 0.7711 \\
CLSpoken1 & 250 & 79.52 & 77.39 (\emph{74.59}) & 0.8171 \\
CLSpoken2 & 300 & 84.65 & 83.91 (\emph{82.14}) & 0.8540 \\
CLSpoken3 & 350 & 74.91 & 70.08 (\emph{66.33}) & 0.8006 \\
\end{tabular}
}
\label{tab:validationbin}
\end{center}
\end{table}

\begin{table}[hb]
\newcommand\T{\rule{0pt}{2.5ex}}       
\caption{Validation performance for TD-TreeLSTM with Vectorial Output}
\begin{center}
\resizebox{.47\textwidth}{!}
{
\begin{tabular}{l|c|c|c|c}
\multicolumn{1}{c|}{\bf{Corpus}} & \bf{Memory Size} & \bf{Accuracy \%} & \bf{Compress. (\emph{Gold}) \%} & \bf{t}\\
\hline \T
CLWritten & 250 & 77.56 & 85.03 (\emph{73.21}) & 0.7075 \\
CLSpoken1 & 200 & 82.66 & 94.12 (\emph{74.59}) & 0.7260\\
CLSpoken2 & 200 & 87.13 & 94.92 (\emph{82.14}) & 0.7998\\
CLSpoken3 & 300 & 77.22 & 79.11 (\emph{66.33}) & 0.7538\\
\end{tabular}
}
\label{tab:validationwe}
\end{center}
\end{table}

Given the validation results, in the following we focus on analyzing the performance of the binary output configuration. Test set assessment has been performed for the best configurations on the validation set: $10$ independent runs with different random initialization of the weights have been performed and results averaged.  Table \ref{tab:test} reports the resulting test set performance: variance is not reported because it is in the order of $10^{-5}$.
\begin{table}[tb]
\newcommand\T{\rule{0pt}{2.5ex}}       
\caption{Test Results for selected configurations of TD-TreeLSTM}
\begin{center}
\resizebox{.47\textwidth}{!}
{
\begin{tabular}{l|c|c|c}
\multicolumn{1}{c|}{\bf{Corpus}} & \bf{Accuracy \%} & \bf{Compress. (\emph{Gold}) \%} & \bf{F1 score \%}\\
\hline \T
CLWritten &  73.58 & 72.37 (\emph{70.41}) & 76.03 \\
CLSpoken1 &  78.44 & 79.50 (\emph{76.45}) & 79.14 \\
CLSpoken2 &  85.05 & 86.27 (\emph{83.83}) & 87.31 \\
CLSpoken3 &  76.18 & 73.40 (\emph{72.04}) & 76.64 \\
\end{tabular}
}
\label{tab:test}
\end{center}
\end{table}

A comparison between our model and state-of-the-art approaches for sequence (LSTM) and tree-based (STSG) compression is reported in Table \ref{tab:comparisons}: our model outperforms both on accuracy/F1 score and compression rate. In particular, our model halves the difference between compression rate and the reference gold compression whilst the accuracy respect to the LSTM-based model is increased by 4\%. F1 score improvements respect to the STSG approach are even more evident (+10\% on CLWritten and +30\% on CLSpoken).
\begin{table}[tb]
\newcommand\T{\rule{0pt}{2.5ex}}       
\caption{Test Performance Comparison with State-of-the-Art}
\begin{center}
\resizebox{.47\textwidth}{!}
{
\begin{tabular}{l|c|c|c}
\multicolumn{1}{c|}{\bf{Model}} &\bf{Accuracy \%} & \bf{Compress. (\emph{Gold}) \%} & \bf{F1 score \%}\\
\hline
\multicolumn{4}{c}{{CLWritten Corpus}}\\
\hline \T
TD-TreeLSTM & 73.58 & 72.37 (\emph{70.41})  & 76.03 \\
LSTM-based\cite{compressLSTM} & \textless70 & 82 (\emph{72}) & - \\
STSG\cite{LapataCompress}&  - & 76.52 (\emph{70.24}) & 66.30 \\
\hline
\multicolumn{4}{c}{{CLSpoken Corpus}}\\
\hline \T
TD-TreeLSTM &  78.89 & 79.72 (\emph{76.82}) & 81.03 \\
LSTM-based\cite{compressLSTM} & \textless75 & - & - \\
STSG\cite{LapataCompress} & - & 82.30 (\emph{76.11}) & 52.02 \\
\end{tabular}
}
\label{tab:comparisons}
\end{center}
\end{table}

\begin{figure}[tb]
\centering
\resizebox{.485\textwidth}{!}
{
\begin{tikzpicture}
    \begin{axis}[
        width  = 0.85*\textwidth,
        height = 8cm,
        major x tick style = transparent,
        ybar=2*\pgflinewidth,
        bar width=10pt,
        ymajorgrids = true,
        ylabel = {Class dimension (\%)},
        xlabel = {Accuracy classes},
        symbolic x coords={30\%-40\%,40\%-50\%,50\%-60\%,60\%-70\%,70\%-80\%,80\%-90\%,90\%-100\%},
        xtick = data,
        scaled y ticks = false,
        ytick distance=5,
        enlarge x limits=0.08,
        ymin=0,
        area legend,
        legend cell align=left,
        legend pos = north west
    ]
        \addplot[style={black,fill=_blue,mark=none}]
            coordinates {(30\%-40\%, 0) (40\%-50\%,2) (50\%-60\%,10) (60\%-70\%, 25) (70\%-80\%,35) (80\%-90\%,18) (90\%-100\%,10)};

        \addplot[style={black,fill=_red,mark=none}]
             coordinates {(30\%-40\%, 1) (40\%-50\%,1) (50\%-60\%,7) (60\%-70\%, 19) (70\%-80\%,28) (80\%-90\%,23) (90\%-100\%,21)};

        \addplot[style={black,fill=_green,mark=none}]
             coordinates {(30\%-40\%, 1) (40\%-50\%,1) (50\%-60\%,1) (60\%-70\%, 8) (70\%-80\%,20) (80\%-90\%,34) (90\%-100\%,35)};

        \addplot[style={black,fill=_yellow,mark=none}]
             coordinates {(30\%-40\%, 1) (40\%-50\%,1) (50\%-60\%,6) (60\%-70\%,22) (70\%-80\%,32) (80\%-90\%,21) (90\%-100\%,17)};

        \legend{CLwritten, CLSpoken1, CLSpoken2, CLSpoken3}
    \end{axis}
\end{tikzpicture}
}
\caption{Test result grouped by accuracy classes for TD-TreeLSTM.}
\label{fig:distr}
\end{figure}
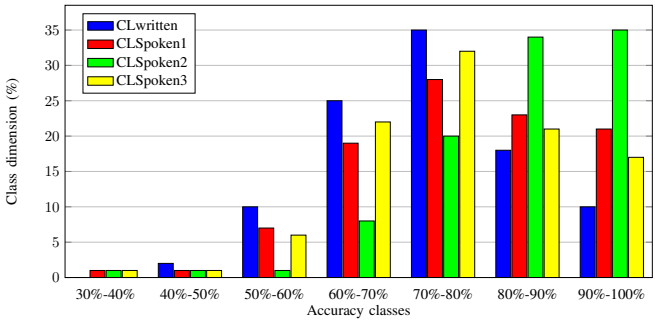

A more detailed view of the precision of the proposed method can be gained by looking at the distribution of the sentences according to accuracy levels. Fig. \ref{fig:distr} shows the number of test set sentences for $7$ accuracy levels:
sentences with an accuracy below 50\% are very few (i.e. less than 10 for each corpus), while on average most corpora sentences fall into the 75\% accuracy class. The CLSpoken2 corpus has a better behavior with most of the sentences falling into the top-accuracy groups.

To appreciate the quality of the generated compressions, Table \ref{tab:best} reports an example of the most accurate compressions (i.e. accuracy\textgreater 90\% excluding the perfect ones to avoid trivial examples). One can see how the common mistakes relate to an excessive deletion of minor elements, e.g. a comma, an article or an adjective. Table \ref{tab:worst} provides examples of worst-case outputs (accuracies\textless 40\%): interestingly, these are typically very short sentences, where missing or adding a deletion has an high impact on sentence accuracy. In this cases, even if the produced output is quite different from the gold compression it is still (mostly) reasonable grammatically and semantically.
\begin{table}[tb]
\newcommand\T{\rule{0pt}{2.5ex}}       
\caption{Examples of best compressions (excluding the perfect ones), for each corpus}
\begin{center}
\begin{tabular}{p{.07\textwidth}|p{.35\textwidth}}
 \multicolumn{1}{c|}{\bf{Corpus}} & \multicolumn{1}{c}{\bf{Sentences}}\\
\hline \T
CLWritten & Input:  When Los Angeles hosted the Olympics in 1932, Kurtz competed in high platform diving.\newline Output: When Los Angeles hosted the Olympics, Kurtz competed in diving.\newline Gold: When Los Angeles hosted the Olympics, Kurtz competed in high diving.\\
\hline \T
CLSpoken1 & Input: As CNN's Jill Dougherty reports, the White House is already trying to calm Netanyahu's critics.\newline Output: White House is trying to calm Netanyahu's critics.\newline Gold: The White House is trying to calm Netanyahu's critics.\\
\hline \T
CLSpoken2 & Input: Direct talks didn't work, and mediators didn't help. \newline Output: Direct talks didn't work and mediators didn't help. \newline Gold: Direct talks didn't work, and mediators didn't help.\\
\hline \T
CLSpoken3 & Input: The FBI has brought in a religious expert. \newline Output: The FBI has brought in religious expert. \newline Gold: The FBI has brought in a religious expert.\\
\end{tabular}
\label{tab:best}
\end{center}
\end{table}
\begin{table}[tb]
\newcommand\T{\rule{0pt}{2.5ex}}       
\caption{Examples of poor sentence compressions for each corpus}
\begin{center}
\begin{tabular}{p{.07\textwidth}|p{.35\textwidth}}
 \multicolumn{1}{c|}{\bf{Corpus}} & \multicolumn{1}{c}{\bf{Sentences}}\\
\hline \T
CLWritten & Input: Mr Salam, 84, a Sunni Muslim,  is the most impressive of Lebanon's dying breed of elder statesmen, but his words could have been those of a much younger man.\newline Output: Mr Salam is the most impressive Lebanon's dying statesmen, but his words could have been those of a younger.\newline Gold: Mr Salam, 84, a Sunni Muslim, is the most impressive of Lebanon's elder statesmen. \\
\hline \T
CLSpoken1 & Input: Ask her whatever you want.\newline Output: Ask her you want.\newline Gold: Ask her whatever.\\
\hline \T
CLSpoken2 &Input: Back to you folks.\newline Output: To folks.\newline Gold: Back to you.\\
\hline \T
CLSpoken3 & Input: And Chicago could take its fourth championship of the decade with one more win.\newline Output: Chicago could take championship with one more win.\newline Gold: Chicago could take its fourth championship of the decade. \\
\end{tabular}
\label{tab:worst}
\end{center}
\end{table}

\section{Concluding Remarks}
We have introduced a deep learning approach to extractive natural language compression by casting it as a restricted form of (parse) tree transduction.
In particular, we proposed the first use of a TD-TreeLSTM model for learning structure-to-substructure transductions by building on application-driven considerations concerning context propagation. We have formulated the claim that a TD approach can help disambiguating the interpretation of the most impacting parts of the parse tree, i.e. the leaves encoding the vocabulary words. We have empirically shown that TD processing allows to exploit such enriched contextual information brought to vocabulary words by the parse tree representation of sentences. The experimental results shows that our solution outperforms the state-of-the-art.

The proposed model can be made more effective in terms of compression performance by working on optimizing the representation of the vocabulary words and syntactic categories. On the one hand, we would like to explore alternative word representations from the NLP literature. On the other hand, we would like to surpass such static representations, introducing a fine-tuning refinement of the label representation by exploiting end-to-end differentiability of the model. Ultimately, we would like to enrich the TD model along the lines of \cite{ordTree}, introducing an additional dependency of the current node from its preceding sibling, defining a form of Parent-Sibling Top-Down TreeLSTM which can be particularly effective in further disambiguating words at the leaves of the parse tree.

\section*{Acknowledgment}
This work has been supported by the Italian Ministry of Education, University, and Research (MIUR) under project SIR 2014 LIST-IT (grant n. RBSI14STDE). The authors gladly acknowledge Dell-EMC and Nvidia for donating the Dell C4130 server and the Nvidia M40 GPUs used to perform the experimental analysis.

\end{document}